\documentclass[journal,comsoc]{IEEEtran}

\usepackage[english]{babel}
\usepackage{graphicx}
\usepackage{epstopdf}
\usepackage{acronym}
\usepackage{balance}
\usepackage{amsmath}
\usepackage{mathtools}
\usepackage{comment}
\usepackage{caption}
\usepackage{enumitem}
\usepackage{adjustbox}
\usepackage{acronym}
\usepackage{url}
\usepackage{breakurl}
\usepackage{array}
\usepackage{pifont}
\usepackage{color}

\acrodef{GNSS}{Global Navigation Satellite System}
\acrodef{SDR}{Software Defined Radio}
\acrodef{IoT}{Internet of Things}
\acrodef{PMU}{Phasor Measurement Units}
\acrodef{GPS}{Global Positioning System}
\acrodef{AI}{Artificial Intelligence}
\acrodef{IMSI}{International Mobile Subscriber Identity}
\acrodef{IMEI}{International Mobile Equipment Identity}
\acrodef{ETA}{Estimated Time of Arrival}
\acrodef{DDoS}{Distributed Denial of Service}
\acrodef{USRP}{Universal Software Radio Peripheral}

\newcolumntype{M}[1]{>{\centering\arraybackslash}m{#1}}

\newcolumntype{P}[1]{>{\centering\arraybackslash}p{#1}}
\newcommand{\cmark}{\ding{51}}%
\newcommand{\xmark}{\ding{55}}%

\begin{document}

\title{Road Traffic Poisoning of Navigation Apps: Threats and Countermeasures}

\author{\IEEEauthorblockN{Simone Raponi\IEEEauthorrefmark{1}, Savio Sciancalepore\IEEEauthorrefmark{2}, Gabriele Oligeri\IEEEauthorrefmark{1}, and Roberto Di Pietro\IEEEauthorrefmark{1}}\\
\IEEEauthorblockA{\IEEEauthorrefmark{1}Division of Information and Computing Technology, \\College of Science and Engineering, \\Hamad Bin Khalifa University, \\ Qatar Foundation \\ Doha, Qatar\\}
\IEEEauthorblockA{\IEEEauthorrefmark{2} Department of Mathematics and Computer Science \\ Eindhoven University of Technology \\  Eindhoven, The Netherlands}}

\graphicspath{{figures/}}

\maketitle

\begin{abstract}
Assisted-navigation applications have a relevant impact on our daily life. However, technological progress in virtualization technologies and Software-Defined Radios recently enabled new attack vectors, namely, \emph{road traffic poisoning}. These attacks open up several dreadful scenarios, which are addressed in this contribution by identifying the associated challenges and proposing innovative countermeasures.
\end{abstract}

\section{Introduction}
\label{sec:intro}

On February 1st, 2020, Simon Weckert transported 99 second-hand smartphones in a handcart to generate virtual traffic jams in Google Maps~\cite{waspost_berlin}. This experiment was pretty successful since he was able to change the traffic state reported by the app from total green (no traffic) to red (congested), despite there were no cars in the surroundings.
From the navigation apps perspective, users can be either \emph{passive actors}, by simply sharing their positions with the platform and let it infer on the current state of the traffic, or \emph{active actors}, actively providing information about the current state of the road traffic. Indeed, the vast majority of assisted navigation applications, such as  Waze, Google Maps, and Apple Maps, to cite a few, allow users to actively report information, signaling events like constructions and traffic incidents.
On the one hand, this feature contributes with useful real-time information to re-route and to provide a more accurate \ac{ETA}. On the other hand, if maliciously exploited, these reporting functionalities can severely affect the road traffic state, for instance, by causing traffic jamming or isolating targeted areas.

Google Maps is currently used by more than one billion monthly active users, representing not only a tool for turn-by-turn navigation, but also an effective way to reduce the \ac{ETA}.
Therefore, injecting massive amounts of fake information (or a few targeted ones) can affect people's daily activities, even their safety. It could also be the case that the above steps are instrumental for the execution of a more complex malicious plan---think about generating traffic jams in specific areas to slow down the intervention of law enforcement officers. Regardless of the motivation, we refer to the previous behavior as \emph{road traffic poisoning}.

The major issue related to the above-introduced vulnerabilities of assisted-navigation apps is represented by the untrusted source of the information those apps rely on~\cite{wang2018_tnw}.
Indeed, road traffic poisoning can be achieved not only by maliciously setting up a large number of devices, but also, and more effectively, by potentially scaling up the attack from a remote and undetected location. For instance, this might be possible by either compromising the positions of a large number of users (i.e., spoofing the \ac{GNSS} signal), or generating a collection of virtual devices with colluding positions and movements.

\ac{GNSS} spoofing is an effective and efficient technique to let a target device estimate a fake position. It can be performed with a cheap \ac{SDR}
and it can affect a significant number of devices in the neighborhood of the \ac{GNSS} spoofer, depending on its transmitting capabilities~\cite{oligeri2019_wisec}. This attack becomes particularly effective when the adversary exploits the presence of many real devices in a crowded place. Indeed, by spoofing their positions, it can affect the traffic status of any place in the world. \\
The above scenario can be further magnified when considering the increasing number of smart devices that are becoming interconnected. Indeed, the number of \ac{IoT} devices is estimated to reach 75.44 billion by 2025 and, being the largest platform of interconnected devices, there have been already cases where a subset is compromised and used to launch \ac{DDoS} attacks. It is just a matter of time when a large number of devices will be exploited and used for road traffic poisoning. Given the number of involved devices, it might be possible to create fake traffic congestion in different areas of several cities, and therefore, countrywide.

However, the most dreadful attack scenario, which is both convenient and scalable from an adversarial point of view, is the one where virtual devices come into play. 
Smartphone virtualization can be easily achieved by tools such as Android Studio, enabling the generation of hundreds, if not thousands, of virtual smartphones reporting fake \ac{GPS} positions. It is worth mentioning that this attack does not involve any physical device, hence being also relatively cheap, it can be launched from any place in the world, and leaves little hope for attribution.

\section{Related Work}
\label{sec:related}

Several studies available in the literature already highlighted some vulnerabilities associated with automatic navigation systems. \textcolor{black}{
For instance, the authors in~\cite{zeng2018_usenix} demonstrated that an automatic navigation system can be cheated by a co-located smart attacker able to intelligently spoof the \ac{GNSS} signal. Specifically, the attack can be carried out in a way to detour the user via a path that has similar crosses and turns of the underlying real path, differing only in a few of them. Similarly, some contributions already identified the security issues related to fake reporting in mobile crowdsourced systems, and pointed out some high-level countermeasures~\cite{restuccia2017_tosn}.
}

As per \ac{GNSS} spoofing, this issue has generated a very active research thread, focused on a few scenarios, including smart grids~\cite{risbud}, turn-by-turn navigation~\cite{wang2018_tnw}, and cyber-physical systems~\cite{wei}. 

Detecting the presence of a \ac{GNSS} spoofer is a challenging task that, during the years, has been carried out mainly by relying on alternative pieces of information. The authors in~\cite{oligeri2019_wisec} suggested exploiting cellular networks beaconing information to implement device localization with an error that is acceptable for many real-life scenarios. The cited solution does not require any intervention on the network infrastructure, while introducing only a minimal overhead to the user device. Similarly, the authors in~\cite{oligeri2020_wisec} used unencrypted broadcast messages emitted by IRIDIUM satellites to cross-check the position computed via standard \ac{GNSS} technologies. The authors in~\cite{Jansen2018} proposed an independent infrastructure to collect (spoofed) positions from airborne traffic and to infer the presence of a \ac{GPS} spoofer. Their solution monitors the air traffic from \ac{GPS}-derived positions advertisements that airborne traffic periodically broadcasts for safety purposes. While not requiring any intervention at both the \ac{GNSS} infrastructure and the device side, the aforementioned solution involves the deployment of a significant number of nodes to collect the \ac{GPS}-derived positions.\\
A \ac{GNSS} receiver clock offset attack is proposed in~\cite{jiang}. The authors showed how to exploit \ac{GNSS} spoofing to introduce a clock offset in the \acp{PMU}. Such an attack introduces an error in the receiver clock offset and, therefore, in the phase error between the voltage and current phase measurements provided by the \ac{PMU}. 

Device cooperation is proposed in~\cite{heng} to detect and mitigate \ac{GPS} spoofing attacks. Each device in the network correlates its position with the one received by other peers belonging to the same network---hence inferring on positioning mismatches. Authors proved that cooperative position authentication can be effectively adopted to detect and mitigate \ac{GPS} spoofing attacks.

The Doppler effect is used to verify the integrity of the \ac{GPS} signal in~\cite{vanmastrigt}. Authors proved that their solution can reliably predict the Doppler shift of \ac{GPS} signal with an accuracy of about 1Hz, highlighting how the Doppler effect can be adopted for \ac{GPS} spoofing detection.

Authors in~\cite{magiera} proposed a spatial processing technique involving an antenna array to measure the phase delay of an actual GPS signal and so inferring on its authenticity. The core idea is rooted in the fact that spoofed signals arrive at the receiver from the same direction, being the spoofer a single transmitting source. Conversely, genuine \ac{GPS} signals arrive from different locations given the nature of the \ac{GPS} satellite constellation.


A summary and comparison of the contributions discussed above are reported in Table~\ref{tab:related}. \textcolor{black}{
It is worth noting that, in principle, all the above strategies can be integrated on commercial mobile and IoT devices to detect an ongoing \ac{GNSS} spoofing attack. Results may be reported to a navigation software to invalidate the attack. 
However, most of them would require either new hardware or dedicated, ad-hoc infrastructure. The high cost of the highlighted requirements makes it difficult for the cited solutions to be massively deployed. 
Moreover, all the above-discussed solutions always assume the target device to be collaborative with the detection system. However, this latter hypothesis is quite restrictive. Indeed, if the device is intentionally carrying out the attack, or it has been compromised, the cited applications can be easily stopped, inhibited, or cheated, thus resulting ineffective.
}

\begin{table*}
\caption{Overview of related work on \ac{GNSS} spoofing detection and derived system requirements}
\centering
\begin{adjustbox}{max width=\textwidth}
\begin{tabular}{|M{1cm}|M{3cm}|M{3cm}|M{3cm}|}
\hline
 \textbf{Ref.} & \textbf{No Auxiliary Ad-Hoc Infrastructure} & \textbf{No Hardware Update/Swap} & \textbf{Effective when the Device is Compromised} \\
\hline
 \cite{oligeri2019_wisec} & \cmark & \cmark & \xmark \\ \hline
 \cite{oligeri2020_wisec} & \cmark & \xmark & \xmark \\ \hline
 \cite{Jansen2018} & \xmark & \xmark & \xmark \\ \hline
 \cite{jiang} & \cmark & \xmark & \xmark \\ \hline
 \cite{heng} & \xmark & \cmark & \xmark \\ \hline
 \cite{vanmastrigt} & \cmark & \xmark & \xmark \\ \hline
 \cite{magiera} & \cmark & \xmark & \xmark \\ \hline
\end{tabular}
\end{adjustbox}
\label{tab:related}
\end{table*}

\textcolor{black}{
Finally, we notice that the contributions in ~\cite{zeng2018_usenix} and ~\cite{restuccia2017_tosn} only cover a single adversarial model, that is the model $A6$. To the best of our knowledge, all the other adversary models ($A1$-$A5$) presented in our manuscript are not covered by any other scientific contribution, and these considerations apply also for the related countermeasures. Such a discussion constitutes our novel contribution to the state of the art, and we are convinced that it could serve as an inspiration for researchers, industry, and additional organizations working on the reliability and dependability of \emph{navigation apps}.
}

\section{Threat Models}
\label{sec:threat_models}

In this section, we depict some possible objectives motivating the adversary to carry out traffic poisoning attacks, as well as the impact of such attacks.
Overall, the following threat models can be identified:
\textcolor{black}{
\begin{itemize}
    \item \textbf{Fake Traffic Generation.} The adversary aims to make a target road appearing as congested, while in reality it is sustaining just regular traffic. This attack can be used to reroute the traffic to alternative roads. To this aim, the adversary can place a significant number of mobile devices to the target road, each one contributing, with its mere presence, to the (fake) congestion. The navigation software will conclude that such a road is congested and it will suggest the users to avoid it, proposing alternative routes. Overall, this strategy will lead the user to follow the path decided by the adversary.
    \item \textbf{Fake Road De-congestion.} This scenario is the dual of the previously introduced one. That is, the adversary aims to make a target road appearing as free while, in reality, it is congested. This attack will eventually drive the victims to congested roads, significantly delaying or blocking their travel, and worsening the congestion. To achieve this goal, the attacker may rely on a significant number of devices, each one reporting regular travel speed on the target road, to overcome the number of vehicles originally reporting congestion phenomena on the same road. Moreover, if placed in the same geographical location of the devices, the attacker might spoof the \ac{GNSS} position of the devices that were properly contributing to have the road being labeled as congested. Once the \ac{GNSS} position is spoofed, they would stop reporting the information related to the road they are currently on, making the attack easier. This will lead the navigation software to (wrongly) consider the road as not-congested, thus justifying its possible inclusion in the optimal path towards other destinations. Overall, this strategy will lead the user to follow the path decided by the adversary, stopping in a congested road.
\end{itemize}
}

Potential attackers may be motivated by different reasons, some of which are reported below.
\begin{itemize}
    \item \textbf{Financial gain.} 
    Let us assume there are two competing shops in a town, both selling similar items. One of the owners may place fake vehicles on the streets adjacent to the competitor's shop, to simulate road congestion, as well as create fake reports, to make people believe the streets have been temporarily closed. This would discourage possible clients to reach the second shop, thus threatening its potential earnings. The attack can easily scale to increasingly sized targets, such as neighborhoods, city blocks, and cities themselves.

    \item \textbf{Terrorism.} This threat may have tragic consequences when it comes to terrorist purposes. Indeed, a terrorist would be able to carry out two types of attacks: mass attacks, and targeted attacks, respectively. In a mass attack, the terrorist might make a congested road look uncongested. By consulting the navigation app, users would observe a road with low (or no) traffic and may consider taking it, inevitably ending up in the queue. In this way the terrorist would have channeled a large number of vehicles onto a small, eventually congested street.
    In a target attack, the terrorist may create fake traffic to congest several streets of a city, thus forcing a specific target to follow a particular path, known in advance by the terrorist, where he could strike the target.
\end{itemize}

\begin{figure}[htbp]
    \centering
    \includegraphics[width=\columnwidth]{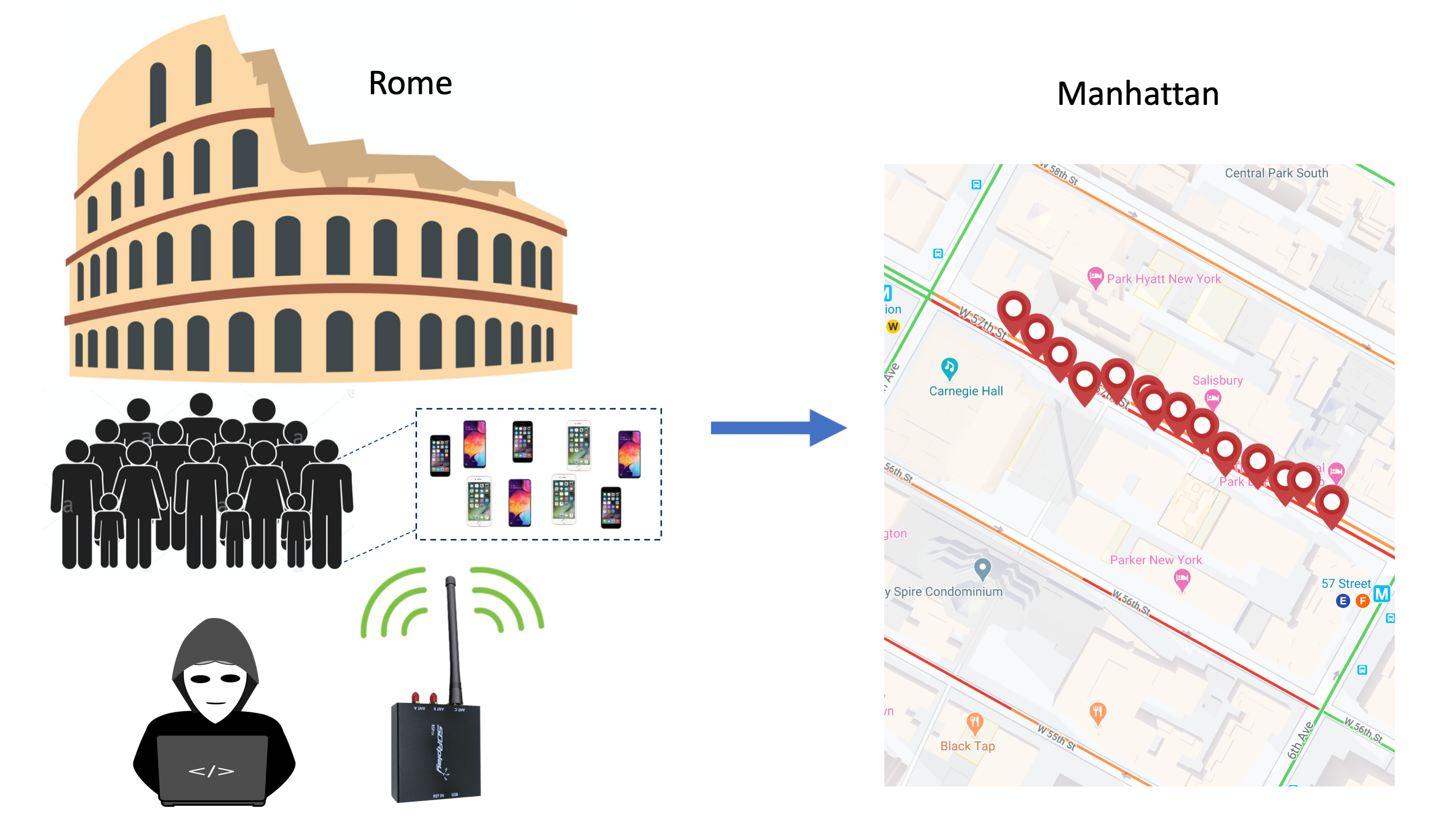}
    \caption{An attacker, equipped with an \ac{SDR}, spoofs the GPS of the mobile devices in a crowded places, geographically teleporting them from their true location (i.e., Rome) to a Manhattan target street, thus making it congested.}
    \label{fig:mass_spoofed_attack}
\end{figure}

\begin{figure}[htbp]
    \centering
    \includegraphics[width=\columnwidth]{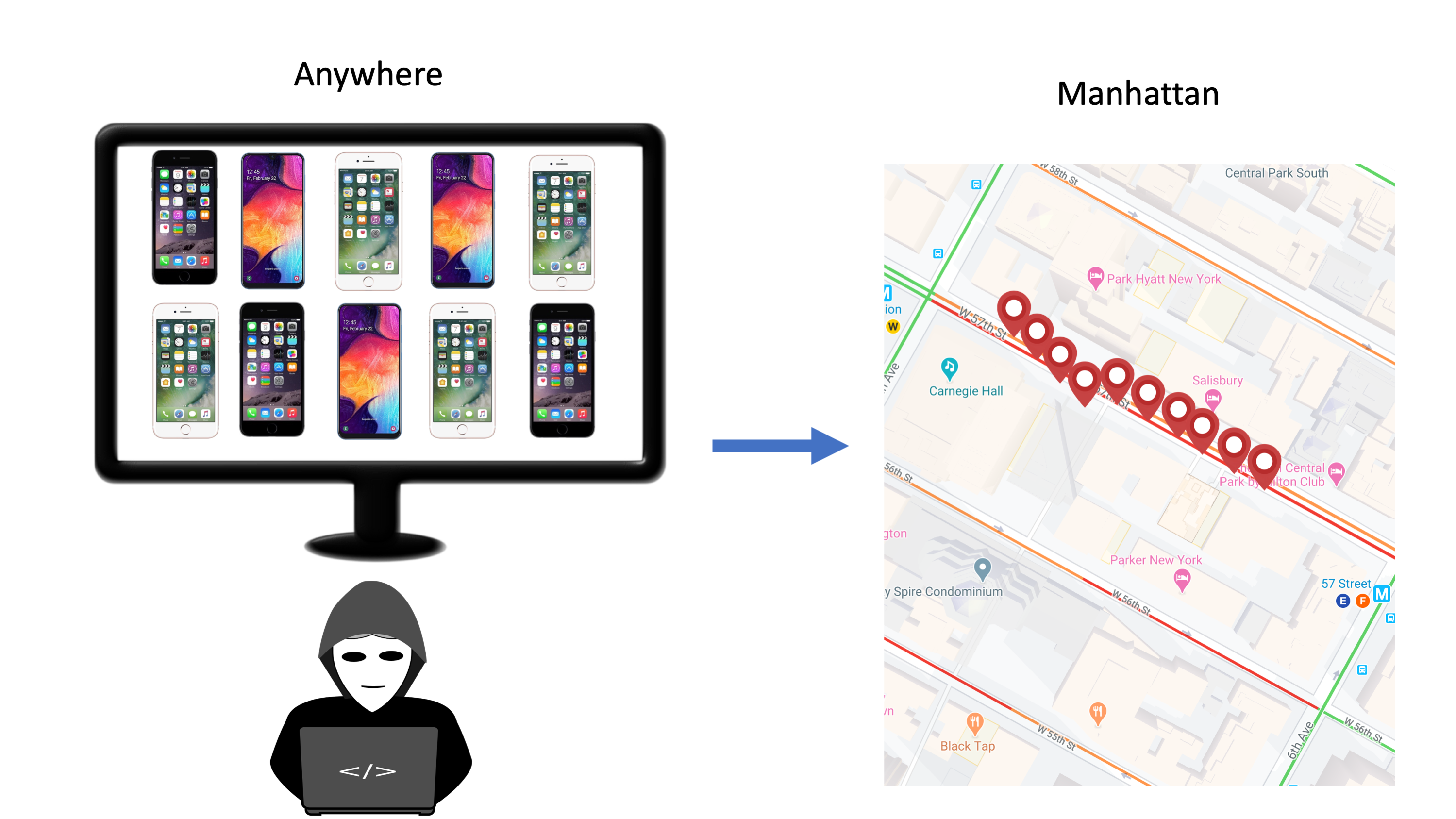}
    \caption{An attacker uses virtual and emulated devices to create fake vehicles to be remotely placed on a target street, thus creating fake road traffic.}
    \label{fig:emulated_spoofing_attack}
\end{figure}

\section{Adversary models}
\label{sec:adversary_models}

Different tools and strategies can be used by the adversary to achieve the goals above described, each characterized by peculiar pros and cons. Some examples are provided below, while a summary of our discussion is synthesized in Table~\ref{tab:adv_models}. \textcolor{black}{It is worth noting that the adversary models discussed in this section involve \emph{external} adversaries, able to carry out \emph{road traffic poisoning} attacks without physically compromising the target vehicle. Therefore, all the attacks requiring the tampering of the integrity of the target vehicle are out of the scope of our investigation.}

\begin{table*}
    \caption{Overview of Adversary Models and Related Aspects.}
    \centering
    \label{tab:adv_models}
    \begin{adjustbox}{max width=0.95\textwidth}
    \begin{tabular}{|M{1cm}|M{2cm}|M{5cm}|M{1.5cm}|M{4cm}|}
        \hline
        \textbf{Adv. Model} & \textbf{Physical vs Remote} & \textbf{Required Tools} & \textbf{Required Skills} &  \textbf{Cost}\\
        \hline
        A1 & Physical & Several mobile devices & Low & Cost of mobile devices, \textit{High}. \\
        \hline
        A2 & Remote & Several mobile devices, laptop, \ac{SDR}, \ac{GPS} spoofing toolkit on laptop & Moderate & Cost of mobile devices, \ac{SDR}, and Laptop, \textit{Very High}. \\
        \hline
        A3 & Remote & laptop, \ac{SDR}, \ac{GPS} spoofing toolkit on laptop & Moderate & Cost of \ac{SDR} and Laptop, \textit{Moderate} \\
        \hline
        A4 & Remote & Emulation Tools, Powerful Workstation & High & Powerful Workstation, \textit{Moderate}. \\
        \hline
        A5 & Remote & Zombies & Very High & \textit{Moderate to High}. \\
        \hline
        A6 & Physical/Remote & Several mobile devices & Low & Cost of mobile devices, \textit{High}.\\
        \hline
    \end{tabular}
    \end{adjustbox}
\end{table*}

\begin{itemize}
    \item \textbf{A1: Real Physically-Located Mobile Devices.} An easy strategy that can be implemented by the adversary requires a significant number of real mobile devices, physically located in the target geographical places. Each device is opportunely set up by the adversary to achieve the desired goal. For instance, if the adversary aims at reporting a congested road, the mobile devices will be placed where needed to emulate road congestion.
    Despite being very simple in practice, such a model requires the adversary to control a significant number of real mobile devices and, possibly, many collaborators. 
    Note that the feasibility of such an adversary model has been recently demonstrated in Berlin, as reported by several media and newspapers~\cite{waspost_berlin}.
    \item \textbf{A2: Real Ad-Hoc Mass Spoofed Mobile Devices.} Another strategy consists of spoofing the \ac{GNSS} position of several dedicated mobile devices that are physically co-located and under the direct control of the adversary. \ac{GNSS} technologies are well-known to be insecure and exposed to easy-to-perform and low-cost \ac{GNSS} spoofing attacks~\cite{oligeri2019_wisec}. Indeed, these attacks can be easily performed by using a cheap \ac{SDR} and a laptop, equipped with a \ac{GNSS} Spoofing Application, such as \emph{GPS-SDR-SIM}~\cite{gpssdrsim}. Thus, the attacker needs to collect many mobile devices and to spoof their geographical location, according to the specific target of the attack. 
    This type of attack can be performed remotely, thus giving up the need for malicious devices to be physically distributed over the target location. However, the attacker should be skilled enough to use \ac{GNSS} spoofing tools.
    \item \textbf{A3: Real Unaware Mass Spoofed Mobile Devices.} Similarly to the previous adversary model, the attacker can spoof the location of mobile devices belonging to unaware, legitimate users, geographically located in the place of interest. \textcolor{black}{However, differently from the adversary model $A2$, the (massively) spoofed mobile devices are not under the direct control/possession of the adversary.}
    To provide a reference example, the attacker can spoof the locations of all the users participating in a big event (e.g., sports events, or concerts) or in a crowded place (e.g., airport, conference hotel). In this way, all the users in the surrounding area will be geographically projected to a target area. Further developments of the attack would make all the users appear moving according to a predefined path, orchestrated by the attacker. An instance of this attack is depicted in Figure~\ref{fig:mass_spoofed_attack}.
    \item \textbf{A4: Emulated Mobile Devices.} Another possible strategy requires the use of emulated mobile devices. By relying on a computational unit with sufficient memory and computational resources (e.g., a laptop or a workstation)---or a set of orchestrated, less powerful units---the adversary can install a mobile device emulation tool, such as \textit{BlueStacks} or \textit{Android Studio Emulator}. Such tools allow launching emulated devices where the attacker has the opportunity to install widely available \ac{GPS} spoofing applications (e.g., FakeGPS Free) to pollute a real \ac{GNSS} position. These applications allow the adversary to geographically (virtually) place the spoofed devices onto a target location. This kind of attack can be performed remotely and using just virtual resources, since neither physical devices nor wireless signals are required. However, the attacker should be skilled enough to master both emulation tools and spoofing applications. This attack is illustrated in Figure~\ref{fig:emulated_spoofing_attack}.
    \item \textbf{A5: Compromised Sparse Devices.} Skilled and motivated adversaries can compromise many real devices (e.g., via trojans or worms), transforming these devices into zombies of a botnet. 
    In this context, the adversary may rely on malicious agents that stealthily install \ac{GNSS} spoofing applications (e.g., the \emph{FakeGPS Free} application previously-mentioned) on the cited devices. 
    When triggered by a remote Command \& Control center, all these devices can be virtually placed to a given target geographical location, thus starting the remote attack. It is worth noting that this strategy extends the number of devices the attacker can rely upon, as even \ac{IoT} devices can be used to launch the attack. While being totally remote and potentially stealthy, the attack requires significant skills from the adversary, as well as time to set up the infecting tools to compose the botnet.
    \item \textbf{A6: Fake Reporting.} Finally, an easy strategy to poison road traffic navigation systems is to report fake data, not corresponding to real situations. This can be achieved by using a massive amount of real mobile devices, reporting all the same conditions, such as the temporary closure of a road, a traffic queue, or an accident. Note that this attack can be either physical, having the devices located at the target roads, or remote, using emulated devices and ad-hoc GNSS spoofing apps. 
    \textcolor{black}{These attacks, well-known in the literature as \emph{sybil attacks}, and whose feasibility has been reported in~\cite{wang2018_tnw}, are possible since the user can affect the state of a target road by reporting fake events, e.g., through the mere  pressing of a button in the application.}
\end{itemize}

\begin{table*}[htbp!]
    \caption{Overview of the proposed countermeasures, pros, and cons.}
    \centering
    \label{tab:countermeasures}
    \begin{adjustbox}{max width=0.95\textwidth}
    \begin{tabular}{|M{2cm}|M{2cm}|M{4.5cm}|M{4.5cm}|}
        \hline
        \textbf{Countermeasure} & \textbf{Implementation Side} & \textbf{Pros} & \textbf{Cons}\\
        \hline
        C1 & Client & Ease of distribution (software update) & Ineffective when the device is willing to attack\\
        \hline
        C2 & Cloud & Ease of distribution and application (software update) & Only increase the difficulty of attack\\
        \hline
        C3 & Cloud & Detection of mass instantaneous moving & Privacy threatening, Computationally-intensive, and time-consuming\\
        \hline
        C4 & Cloud & No fake emulated devices & Limiting User Devices\\
        \hline
        C5 & Cloud & Lightweight and Easy to apply (software update) & Risks of Collusion\\
        \hline
        C6 & Cloud & Effective & High Costs, Privacy Risks\\
        \hline
    \end{tabular}
    \end{adjustbox}
\end{table*}

\section{Enabling Tools}
\label{sec:enabling_tools}

In this section, we describe the rationale and the setup of the operation tools used in this study to launch road traffic poisoning attacks.
We focus on two main scenarios: the in-loco fake traffic attack (S1), and the virtual remote traffic attack (S2), respectively.

\textit{S1: In-Loco Fake Traffic Attack.} Consistently with the Adversary Model A2, in this scenario, the attacker has to be in the same geographical location as the victims. The attack requires the following enabling tools.

\begin{itemize}
    \item \textbf{Spoofing Hardware Device.} The hardware required to spoof a \ac{GNSS} signal mainly consists of an \ac{SDR}, such as HackRF, LimeSDR, or Ettus \ac{USRP}, to cite a few. \ac{SDR}s allow implementing via software a set of radio components that are traditionally implemented via hardware.
    \item \textbf{Software Development Toolkit.} One of the most commonly adopted platform to deploy proof-of-concept radio applications is GNURadio, defined as an open-source software development toolkit featuring a significant number of already implemented processing blocks.
    \item \textbf{Spoofing Module.} One of the most adopted software tools to spoof a GNSS signal is \emph{GPS-SDR-SIM}, which is an open-source software-defined \ac{GPS} signal simulator featuring several spoofing options such as static locations, time synchronization, and paths with different speeds.
\end{itemize}
    
\textit{S2: Virtual Remote Traffic Attack.} Consistently with the Adversary Model A4, in this scenario, the attacker has the opportunity to perform a remote attack to change the road traffic conditions anywhere in the world. The attack requires the following enabling tools.

\begin{itemize}
    \item \textbf{Mobile Terminal Emulator.} The attacker relies on tools enabling Android applications to run on PCs, regardless of the host operating systems. Tools of this type include Android Studio Emulator, BlueStacks, and Genymotion Android Emulator, to name a few.
    \item \textbf{\ac{GPS} Faker.} Once emulated the Android devices, the attacker should exploit apps, web services, and other means to spoof their \ac{GNSS} locations. The reference Android market, i.e., Google Play Store, boasts many working applications, such as ``Fake GPS GO Location Spoofer Free'', that allows being instantly geographically teleported anywhere in the world, as well as to set specific routes the fake (i.e., emulated) person will run across. Such a service is installed by default on some emulators. For instance, the BlueStacks emulator includes an application, namely ``Location Provider'', that allows to statically set a geographical location to be teleported to, and such application can be easily installed in other emulators.
    \item \textbf{Support PC.} To run the emulated devices, the adversary needs a PC to support virtualization. The performance of the PC (i.e., the amount of memory and computational capabilities) will dictate the number of devices that can be emulated. It is worth noting that the adversary may rely on powerful servers, as well as on virtual machines/containers from the cloud (i.e., Amazon Web Services, Microsoft Azure, Docker, to name a few) to carry out the attack, having the opportunity to disseminate a huge number of fake devices and provoke heavy and potentially dangerous services outage.
\end{itemize}

\section{Countermeasures}
\label{sec:countermeasures}

\textcolor{black}{
In this section, we describe some countermeasures to the threats and attacks above described. A summary of the following discussion is synthesized in Table~\ref{tab:countermeasures}. We refer to each of the identified countermeasures with an ordinal number, as indicated in Table~\ref{tab:countermeasures}, while the attack scenario tackled by each countermeasure is reported in the related description. \\
Overall, the strategies to mitigate the above attacks could be distinguished in client-side countermeasures and cloud-side countermeasures.
On the client-side, a countermeasure to prevent the \emph{road traffic poisoning} attack consists in detecting a GNSS spoofing attack. Specifically, a possible technique to avoid GNSS spoofing implies the presence of benign devices, that are not fully controlled by the adversary and do not participate voluntarily in the attacks. This is the case for the Adversary Model A3. In this scenario, an effective countermeasure consists of applying standard GNSS spoofing detection and mitigation techniques to protect such devices from being part of the attack (C1). For instance, as described by the authors in~\cite{oligeri2019_wisec}, crowdsourced information originated from the mobile cellular network, or any other communication technology such as ADS-B, can be used to detect the \ac{GNSS} spoofing attack~\cite{Jansen2018}. In summary, each vehicle can compute an independent location estimate by only relying on information emitted by dedicated opportunistic sources, such as the mobile cellular network. Despite these location estimates are affected by significant errors and they cannot be used for localization purposes, they can be still useful for cross-checking the information obtained via the regular \ac{GNSS} technologies, thus detecting potential inconsistencies.
However, if the attacker owns the attacking devices, or it has physically compromised the devices by taking full control, the above-described techniques could be made ineffective.
The good news is that, for the cases: A1, A2, A4, A5, and A6 discussed in Section~\ref{sec:adversary_models}, the attacks can be effectively detected, though only by the service provider, i.e. cloud-side. 
Different techniques, listed in the following, can be adopted in this case to either mitigate or, at least, make the attack more difficult. Note that the strategies listed below are \emph{attack detection} strategies, that apply on the server after the injection of the \emph{poisoned} traffic information. \\
\begin{itemize}
    \item \textbf{C2: Crowdsourcing.} The software application running on the devices can gather heterogeneous information in addition to the \ac{GNSS} location only. These pieces of information include, but are not limited to, the instantaneous readings of the accelerometer, the magnetometer, the information about the base stations of the cellular network, and the current IP address. Then, the automatic navigation system servers on the cloud could evaluate and verify the consistency of these pieces of information. While this strategy does not provide a $100\%$ guarantee of detecting the attacks (all the above-reported readings can be falsified and artfully tuned by the adversary, as shown in~\cite{trippel2017_esp}), it significantly increases the difficulty for the attack to succeed, forcing the adversary to coordinate several sources to be effective. 
    \item \textbf{C3: Temporal Analysis.} Using the information available in the cloud, the service provider can adopt \ac{AI} techniques to analyze the behavior of users throughout time, as well as the evolution of the situation for a specific road over time. 
    For instance, the service provider can detect an attack if a road suddenly (or in a very short time frame) becomes congested shortly after being used by a few users. Similarly, an immediate and mass movement of a group of users from the same place to a far location could be easily and quickly detected. 
    \item \textbf{C4: Multiple-Factor Authentication on Mobile Devices with unique value pairs.} To thwart the adversary models A4 and A5 based on emulated devices, a simple countermeasure might consist of verifying that each device reporting information in the navigation system is associated exactly to a single physical user. A method to achieve this objective could rely on multi-factor authentication, leveraging unique \acp{IMSI} and \acp{IMEI}, verifiable by the manufacturer. 
    While being quite effective in reducing the extent of the attack, such countermeasure would require coordination between multiple independent companies, also sharing potentially private data. Thus, it would require significant overhead, time, and resources by the involved companies.  
    \item \textbf{C5: Trust Mechanisms.} An alternative countermeasure could consist of building up a trust logic, where each device injecting information in the system is characterized by a \emph{trust score}. This score could change dynamically, based on the feedback that users provide. Therefore, only when the device injecting information is trusted enough, the specific information can be classified as trusted and used for navigation purposes. Although improving the overall difficulty and time required to attack the system, this countermeasure can be circumvented if the attacker owns a significant number of devices. Indeed, these devices might start colluding and improving each other's trust. Without loss of generality, traditional anti-collusion techniques in trust systems can be used to enhance the robustness of the system.
    Note that if the devices are spoofed by the attacker, this logic would not be effective.
    \item \textbf{C6: Smart Cities Assistance.} By definition, a Smart City is an urban area that relies on smart devices (including, but not limited to, \ac{IoT}) to collect data and to suggest consequent actions. 
    The insights gained from these data will drive every decision finalized to efficiently manage (i.e., optimize) resources, services, and assets. Privacy-preserving measurements could be collected to understand the extent of road traffic on the streets within a city. For instance, smart traffic lights, as well as smart street lamps and smart axle counters, may provide publicly accessible information about the traffic without revealing any information about the vehicles. The automatic navigation system servers might cross-correlate these pieces of information with the user reports to achieve more authoritative knowledge and provide accurate and effective route recommendations. While being quite promising, this solution may be made ineffective if some auxiliary \ac{IoT} devices are physically tampered or hacked.
\end{itemize}
}

\section{Conclusion and Outlook}
\label{sec:conclusion}

\textcolor{black}{
In this paper, we have discussed the objectives, scenarios, models, tools,  and attack vectors related to the {\em road poisoning attack} affecting the navigation apps currently available on the market. We have also discussed potential novel countermeasures to such  disruptive attacks.
Indeed, navigation apps strongly rely on user-provided information---either actively or passively. In the  former case information are sent with the active involvement of the user, while in the latter case information are  collected by the app, for instance in terms of user's position and speed. 
The current literature mainly focuses on \emph{sybil attacks}, aiming at corrupting the information reported to navigation apps via \emph{fake reporting}. However, several other attack vectors can be used to achieve \emph{road traffic poisoning}, relying on GNSS spoofing and software virtualization tools.\\
Overall, we have discussed how the aforementioned information can be corrupted and used for launching attacks that can scale up to hundreds, or even thousands, of (fake) users. Moreover, we have provided a classification of the possible attacks based on the involved resources. Finally, we have discussed possible countermeasures and their limitations, also highlighting related open issues.\\
Despite this paper highlights the feasibility of these attacks only, the models, tools, attack vectors, and countermeasures discussed in the paper, other than being interesting on their own, have also the potential to pave the way for future work in the area, both in the academia and the industry domain.
}

\bibliographystyle{IEEEtran}
\bibliography{biblio}

\end{document}